\documentclass{article}
\usepackage{achemso}
\usepackage[utf8]{inputenc}
\usepackage{textcomp} 
\usepackage{graphicx}
\usepackage{ulem}
\usepackage[dvipsnames]{xcolor}
\usepackage{amssymb}
\usepackage{amsmath}
\usepackage[a4paper, portrait, margin=1in]{geometry}
\usepackage{breqn} 

\title{A new signature for strong light-matter coupling using spectroscopic ellipsometry}
\author{Philip A. Thomas\footnote{p.thomas2@exeter.ac.uk}, Wai Jue Tan, Henry A. Fernandez\footnote{Current address: Department of Electronics and Nanoengineering, Aalto University, Tietotie 3, FI-00076 Espoo, Finland}, \\ and William L. Barnes\footnote{The authors thank Vasyl G. Kravets, Gregory H. Auton and Alexander N. Grigorenko for providing the surface lattice resonance/guided mode structure.
P.A.T. and W.L.B. acknowledge the support of the European Research Council through Project Photmat (ERC-2016-AdG-742222: www.photmat.eu).
W.J.T. and H.A.F. acknowledge financial support from the Engineering and Physical Sciences Research Council (EPSRC) of the United Kingdom via the EPSRC Centre for Doctoral Training in Metamaterials (Grant No. EP/L015331/1).
Data in support of our findings are available at: https://ore.exeter.ac.uk.uk/repository/handle/XXX} \\ \\ \small{Department of Physics and Astronomy, University of Exeter,} \\ \small{Exeter, EX4 4QL, United Kingdom}}
\date{}

\begin{document}

\maketitle

\begin{abstract}
Light-matter interactions can occur when an ensemble of molecular resonators is placed in a confined electromagnetic field.
In the strong coupling regime the rapid exchange of energy between the molecules and the electromagnetic field results in the emergence of hybrid light-matter states called polaritons.
Multiple criteria exist to define the strong coupling regime, usually by comparing the splitting of the polariton bands with the linewidths of the uncoupled modes.
Here we highlight the limitations of these criteria and study  strong coupling using spectroscopic ellipsometry, a commonly used optical characterisation technique.
We identify a new signature of strong coupling in ellipsometric phase spectra. 
Combining ellipsometric amplitude and phase spectra yields a distinct topological feature that we suggest could serve as a new criterion for strong coupling.
Our results introduce the idea of ellipsometric topology and could provide further insight into the transition from the weak to strong coupling regime.
\end{abstract}

\section*{Keywords}
strong coupling, ellipsometry, polaritons, optical microcavities, optical phase response, Rabi splitting

\section{Introduction}

Light-matter interactions can occur when an ensemble of molecular resonators is placed in a confined electromagnetic field.
If the field and resonators have similar excitation energies and the coupling strength between them exceeds the mean of their decay rates, the energy levels of the confined field mode and the resonator can be modified: they are \textit{strongly coupled}\cite{khitrova2006vacuum, torma2014strong}.
The characteristic feature of strong coupling is the formation of two hybrid states known as the upper and lower polariton bands\cite{tame2013quantum}.
Confined electromagnetic fields can be generated by optical microcavities\cite{weisbuch1992observation, lidzey1998strong, schwartz2011reversible} or surface plasmons\cite{bellessa2004strong, fofang2008plexcitonic}; resonances can be provided by organic molecules\cite{lidzey1998strong, bellessa2004strong, fofang2008plexcitonic, schwartz2011reversible, shalabney2015coherent, long2015coherent}. 
The potential of strong coupling to control light-matter interactions is far ranging, with applications identified in the areas of quantum information\cite{romero2012ultrafast,stassi2018long}, polaritonic chemistry\cite{feist2017polaritonic} and lasing\cite{christopoulos2007room}, among others.

Strongly coupled systems are usually characterised by an intensity measurement (such as reflectivity\cite{lidzey1998strong}, extinction\cite{fofang2008plexcitonic}, transmission\cite{schwartz2011reversible} or luminescence\cite{bellessa2004strong}) which is used to create a dispersion plot (energy versus incident angle of light $\theta$ or wavevector $k_{//} = \frac{2\pi}{\lambda} \sin(\theta)$, where $\lambda$ is wavelength of incident light; see figure \ref{fig:expt}a for an example). 
The signature of strong coupling observed in these plots is an anticrossing of the confined electromagnetic mode and the material resonance. 
The Rabi splitting, $\Omega$, is the minimum energy difference between the two modes. 
Multiple criteria for strong coupling exist and are usually defined by comparing the linewidths of the uncoupled resonances with $\Omega$ \cite{torma2014strong}.
These criteria predict the transition from weak to strong coupling at very different values of $\Omega$.

The combined study of the amplitude and phase response of an optical system can provide insight that is not possible from intensity measurements alone.
By analysing the amplitude and phase response of plasmon antenna array etalons, Berkhout and Koenderink\cite{berkhout2019perfect} showed that points of perfect absorption in such structures are topologically protected.
Kravets \textit{et al.}\cite{kravets2013singular} showed that the phase response of plasmonic nanostructures around points of perfect absorption can be used in single molecule detection.
To the best of our knowledge no experiments have studied the phase response in molecular strong coupling.

In this work we use spectroscopic ellipsometry to study the combined amplitude and phase response of strongly coupled resonances. 
We characterise the strong coupling of optical microcavities with organic molecules and that of surface lattice resonances with waveguide modes. 
We observe the transition from weak to strong coupling using the ellipsometric phase shift and identify a candidate signature of strong coupling. 
Combining amplitude and phase data shows that the optical response of the system undergoes a change in topology during the transition from weak to strong coupling.
We compare this transition point with the existing criteria for strong coupling.
Our results suggest a new criterion for strong coupling, free of the limitations of existing strong coupling criteria, and reveal a new way to study the topology of optical systems.

\section{Results}


\begin{figure}
\includegraphics[scale=0.7]{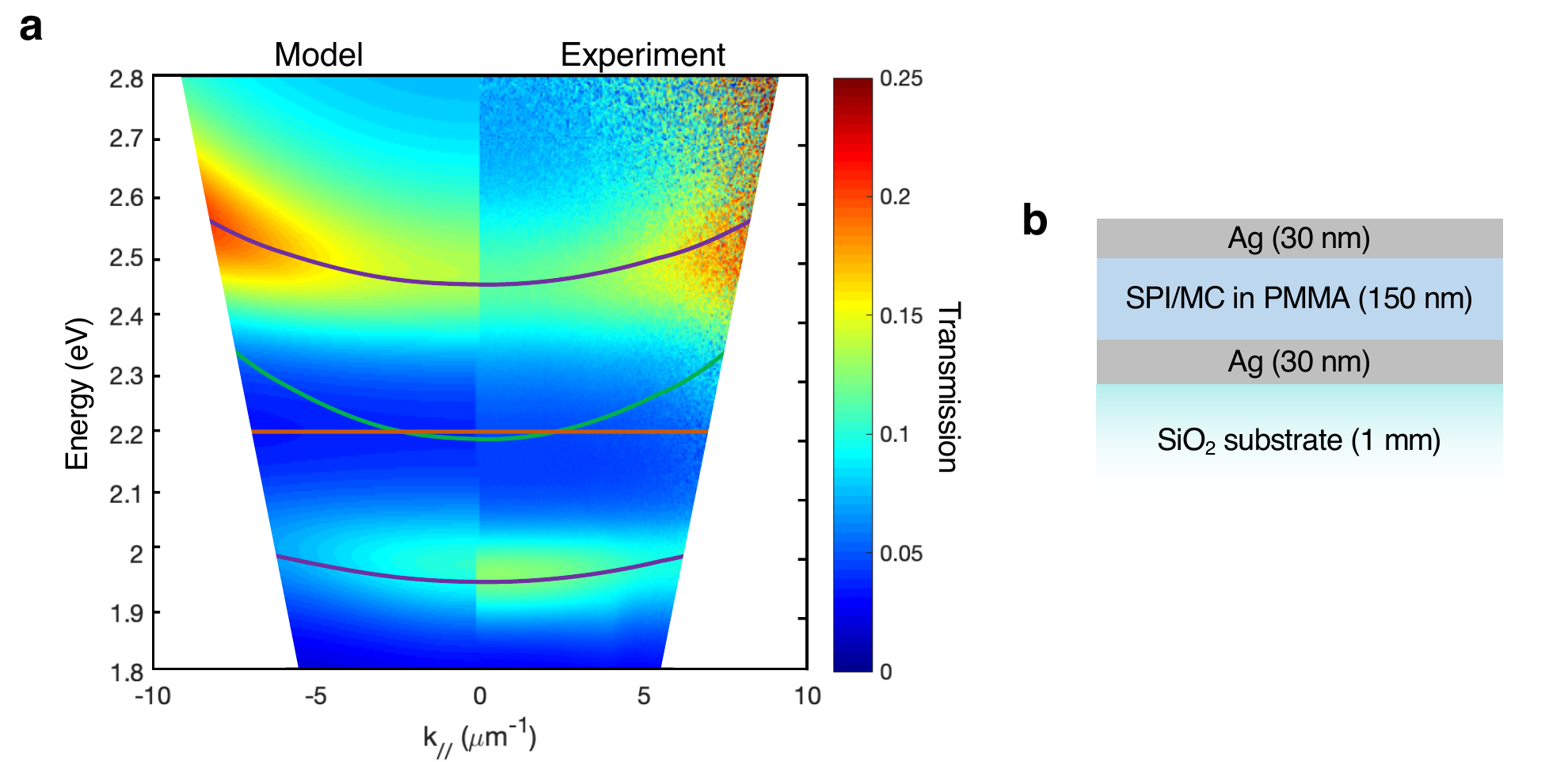}
\centering
\caption{\textbf{Strong coupling between optical microcavity mode and organic molecules.} 
(a) Typical dispersion plot of an optical cavity mode coupled to organic molecules (here merocyanine). 
Data plotted for negative $k_{//}$ were calculated using transfer matrix method (merocyanin was modelled using a Lorentz oscillator).
Data plotted for positive $k_{//}$ were obtained using Fourier transmission spectroscopy (Supplementary Figure S1).
Experimental transmission values have been scaled up by $50\%$ to match calculated data, compensating for a drop in intensity due to scattering.
The green and orange lines indicate the positions of the uncoupled cavity and molecular resonances, respectively; the purple lines show the calculated positions of the upper and lower polariton bands.
(b) Sample design for strong coupling experiment showing a dye-doped polymer matrix between two silver mirrors.}
\label{fig:expt}
\end{figure}

We studied strong coupling between organic molecular resonances and optical cavity modes. 
(See Supplementary Information for fabrication details.)
The microcavity design is illustrated in figure \ref{fig:expt}b: it consisted of two silver mirrors (each of thickness 30 nm) separated by a PMMA (polymethyl methacrylate) dielectric spacer layer (thickness 150 nm).
Embedded in the PMMA layer are spyropyran (SPI) molecules (1',3'-dihydro-1',3',3'-trimethyl-6-nitrospiro[2H-1-benzopyran-2, 2'-(2H)-indole]). 
SPI is a transparent photochromic molecule: after exposure to ultraviolet radiation it undergoes photoisomerisation and is converted to merocyanine (MC)\cite{berkovic2000spiropyrans} with an optical transition at 2.2 eV (Supplementary Figure S2). 
The cavity thickness was chosen so that the first-order cavity resonance occurred at 2.2 eV for light incident at $\theta = 60^{\circ}$. 
Exposing the cavity to ultraviolet radiation converts SPI to MC and the first-order cavity mode couples to the molecular resonance of MC.
This allows for observation of the transition from weak to strong coupling\cite{schwartz2011reversible}. 

\begin{figure}
\includegraphics[scale=0.7]{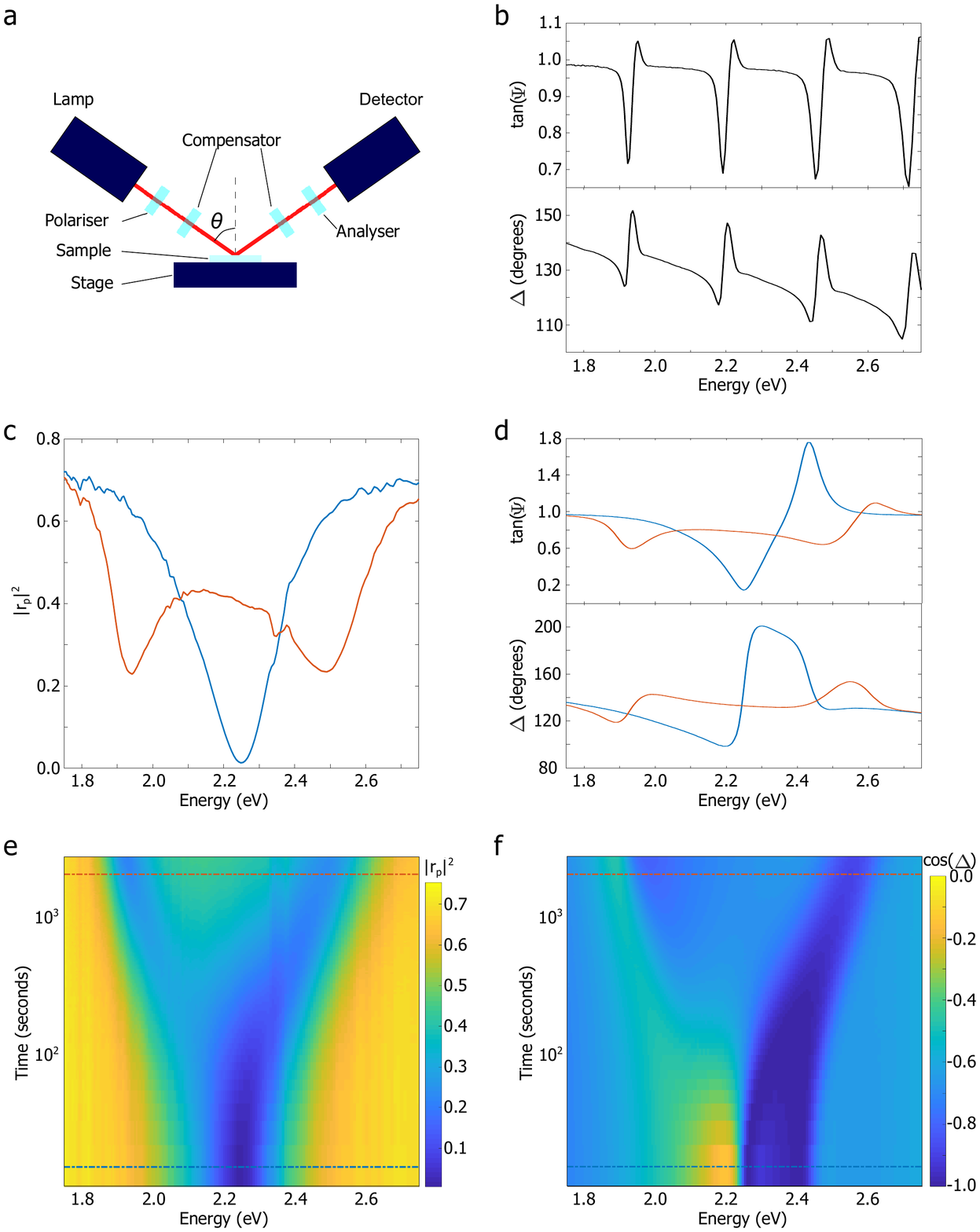}
\centering
\caption{\textbf{Transition from weak to strong coupling regime observed with intensity and phase measurements.} 
(a) Schematic of spectroscopic ellipsometer. 
$\theta=60^{\circ}$ for all measurements.
(b) $\tan(\Psi)$ and $\Delta$ for a series of uncoupled cavity modes in a $\sim 2$ µm-thick PMMA microcavity. 
(c-d) Initial (blue) and final (orange) measurements of SPI/MC microcavity made with (c) $R_p$ and (d) $\tan(\Psi)$ (top) and $\Delta$ (bottom) measurements showing the change from a single uncoupled cavity mode to strongly coupled MC/cavity modes. 
The transition from weak to strong coupling as a function of time is shown with $R_p$ data in (e) and with $\cos(\Delta)$ data in figure (f). The dashed blue and orange lines shows the positions in time from which the data in (c)-(d) were taken.}
\label{fig:time}
\end{figure}


All samples were characterised using spectroscopic ellipsometry (figure \ref{fig:time}a), which measures the complex reflection ratio $\rho$ in terms of the parameters $\Psi$ and $\Delta$:
\begin{equation}
    \rho = \frac{r_p}{r_s} = \tan(\Psi) e^{i\Delta}.
\end{equation}
\noindent $r_p$ and $r_s$ are the Fresnel reflection (amplitude) coefficients for p- and s-polarised light, respectively; 
$\tan(\Psi)$ is the amplitude of $\rho$ and provides the ratio of $r_p$ and $r_s$; whilst $\Delta$ is the difference in the phase shifts undergone by p- and s-polarised light upon reflection. 
(Further details in Supplementary Information.)
The dominant use of spectroscopic ellipsometry is in determining the thickness and optical constants of thin films\cite{tompkins2005handbook, greef1986polar}.

The ellipsometric response of a multimode Ag/PMMA/Ag microcavity (thickness $\sim 2$ µm) at $\theta=60^{\circ}$ is shown in figure \ref{fig:time}b.
Since these measurements were made at an oblique incident angle, the cavity resonances occur at different energies for p- and s-polarised light.
In $\tan(\Psi)$, a resonance occurs when $r_p < r_s$ ($\tan(\Psi)<1$) and also when $r_s < r_p$ ($\tan(\Psi)>1$).
$\Delta$ is the difference between the phase change experienced by p- and s-polarised light: a cavity resonance will cause a characteristic modulation in $\Delta$.

\begin{figure}
\includegraphics[scale=0.4]{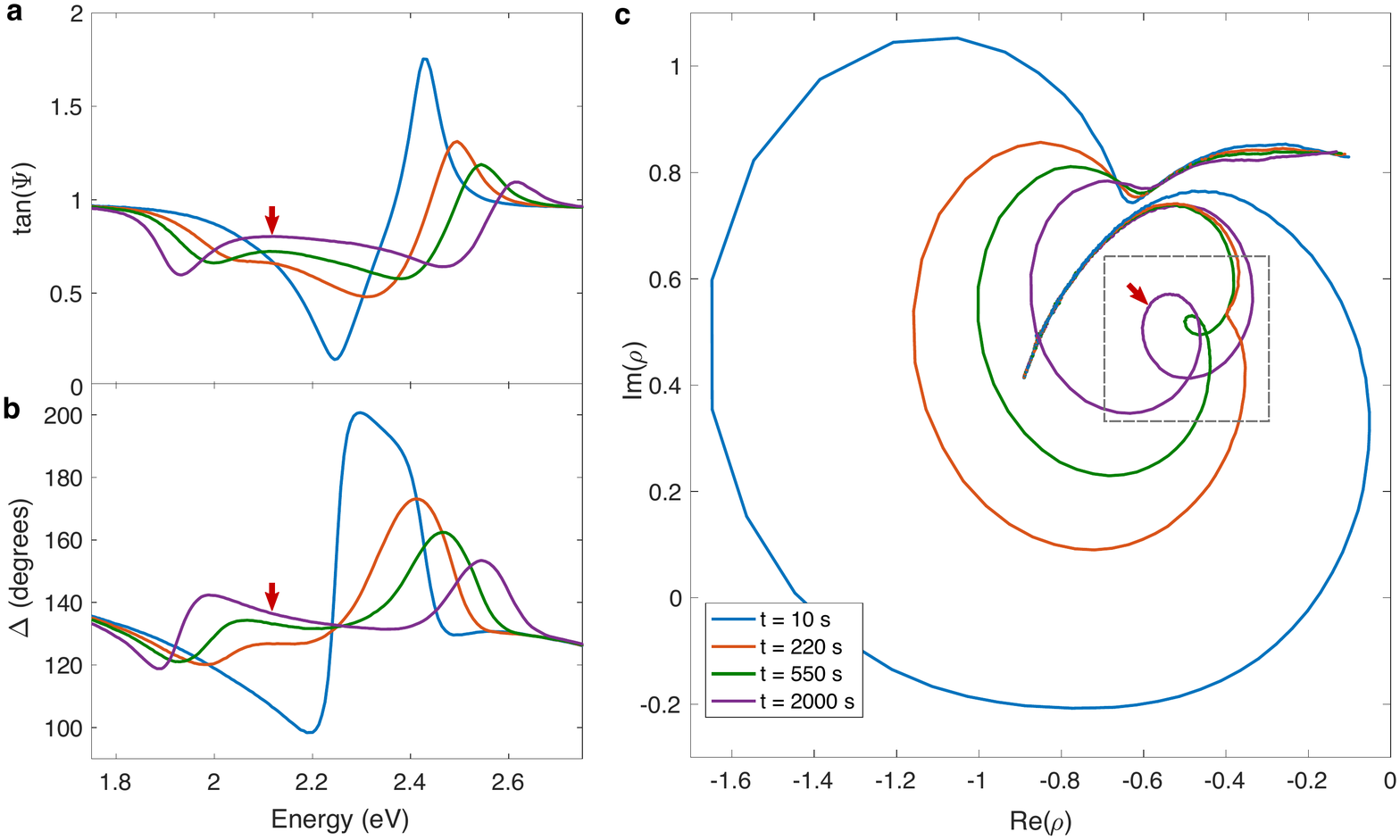}
\centering
\caption{\textbf{Evolution of $\rho$ through transition from weak to strong coupling.}
All measurements made at $\theta=60^{\circ}$.
(a), (b) and (c) show the measured ellipsometric parameters $\tan(\Psi)$, $\Delta$ and $\rho$, respectively, for the SPI/MC microcavity at the times $t=10$ s (blue curve), $t= 220$ s (orange curve), $t=550$ s (green curve) and $t=2000$ s (purple curve).
The critical region of interest in (c) is indicated by the dashed grey box.
The point with maximum $\tan(\Psi)$ between the two polariton minima in the final (purple) dataset is indicated by the red arrows.
}
\label{fig:rho}
\end{figure}

We exposed the SPI microcavity to UV irradiation and measured the change in $\rho$ (figure \ref{fig:rho}), and its derived values ($R_p=|r_p|^2$, $\tan(\Psi)$ and $\Delta$; figure \ref{fig:time}), as SPI underwent conversion to MC. 
All measurements were taken at $\theta=60^{\circ}$. 

Figure \ref{fig:time}c shows the $R_p$ spectrum before and after the SPI microcavity was exposed to ultraviolet radiation.
The MC resonance at 2.21 eV couples to the cavity mode at 2.24 eV; the maximum Rabi splitting observed was $(574 \pm 103)$ meV. 

Figure \ref{fig:time}e shows the time evolution of $R_p$ spectra of the microcavity as SPI is converted to MC. 
The high time resolution of our measurements (one scan every 11 seconds over a total acquisition time of 45 minutes) demonstrates a clear transition of the reflection spectrum from an uncoupled to a coupled state.
After the first hundred seconds of ultraviolet exposure the cavity resonance splits into two and the rate of splitting slows down exponentially. 
As $\Omega$ is directly proportional to $\sqrt{N/V}$ (where $N$ is the number of MC molecules in the cavity and $V$ is the cavity volume)\cite{torma2014strong}, this implies that the conversion of SPI to MC molecules follows an exponential relationship with time.


We plot $\Delta$ in figure \ref{fig:time}d (initial and final state) and \ref{fig:time}f (change with time; cos($\Delta$) has been plotted to improve contrast).
In figure \ref{fig:time}c ($R_p$) the upper and lower polariton bands are, like the original cavity mode, approximately Lorentzian in form. In contrast, the phase signatures of the upper and lower polariton bands in figure \ref{fig:time}d,f have different forms, as if a point of inflection has been added to the centre of the original phase response.
This differs from the phase response of the multimode cavity in figure \ref{fig:time}b which shows closely spaced but uncoupled cavity modes.
Since the MC molecular resonance does not change with $\theta$ and is not polarisation-dependent, the splitting of the asymmetric $\Delta$ response of the microcavity shows that the properties of the original cavity mode have been inherited by the upper and lower polariton bands.
This suggests that phase measurements can distinguish between coupled resonances and uncoupled but closely spaced resonances in a way which is not possible using intensity measurements.


The parameters $\tan(\Psi)$ and $\Delta$ are plotted in figures \ref{fig:rho}a-b.
Four datasets are plotted in these figures: the initial SPI microcavity (blue lines, $t=10$ s); an intermediate point where some SPI has been converted to MC but not sufficiently for strong coupling (orange lines, $t=220$ s); a point at which the resonance is split in both $\tan(\Psi)$ and $\Delta$ (green line, $t=550$ s); and the final strongly coupled MC microcavity (purple line, $t=2000$ s).
These datasets are used to plot $\rho$ in the complex plane as a function of energy (from 1.1 eV - 3.5 eV) in figure \ref{fig:rho}c. 
For a simple silver surface (Supplementary Figure S3a-b) $\rho$ traces out an arc from $E = 1.5$ eV, $\rho \approx -0.8 + 0.55i$ to $E = 3.5$ eV, $\rho \approx -0.1 + 0.85i$. This arc, present in all curves in figure \ref{fig:rho}c, results from the optical response of Ag as it changes from a mirror-like response at lower energies (perfect reflection occurs at $\rho=-1$) towards interband transitions at around 3.9 eV\cite{ehrenreich1962optical}.

The changes in both $\tan(\Psi)$ and $\Delta$ associated with a cavity resonance observed at $\theta \neq 0$ combine to add a loop that breaks the Ag arc in $\rho$.
This loop, representing the first-order microcavity resonance, appears along the Ag arc at $\rho \approx -0.65 + 0.75i$ (see Supplementary Figure S3c-d).

The changes in both $\tan(\Psi)$ and $\Delta$ associated with the first-order cavity resonance combine to add a loop that appears along the Ag arc at $\rho \approx -0.65 + 0.75i$ (see Supplementary Figure S3c-d).


As SPI is converted into MC the area enclosed by the cavity resonance in $\rho$ reduces, corresponding to a decrease in the strength of the cavity resonance. 
As MC is created a ``kink'' appears at $\rho \approx -0.40 + 0.55i$ and grows on the side of the cavity loop (orange curves). 
Figure \ref{fig:rho}a shows this is a change from a single resonance to two resonances that are not yet fully distinct.
Between the orange and green curves two resonances become observable in $\tan(\Psi)$ and the point of inflection in $\Delta$ evolves into the local minimum observed in figure \ref{fig:time}d,f.
In figure \ref{fig:rho}c this corresponds to the dimple evolving into a secondary loop inside the original resonance loop (a change in ellipsometric topology).
The point of maximum $\tan(\Psi)$ between the two polariton minima in the final dataset is indicated by red arrows, showing that here the secondary loop lies between the positions of the two polaritons on the primary loop.
In contrast, plotting uncoupled resonances in $\rho$ gives one independent loop per uncoupled resonance. These loops can overlap but do not form secondary loops; see $\rho$ plotted for a multimode cavity in Supplementary Figure S3e-f.


\begin{figure}
\includegraphics[scale=0.4]{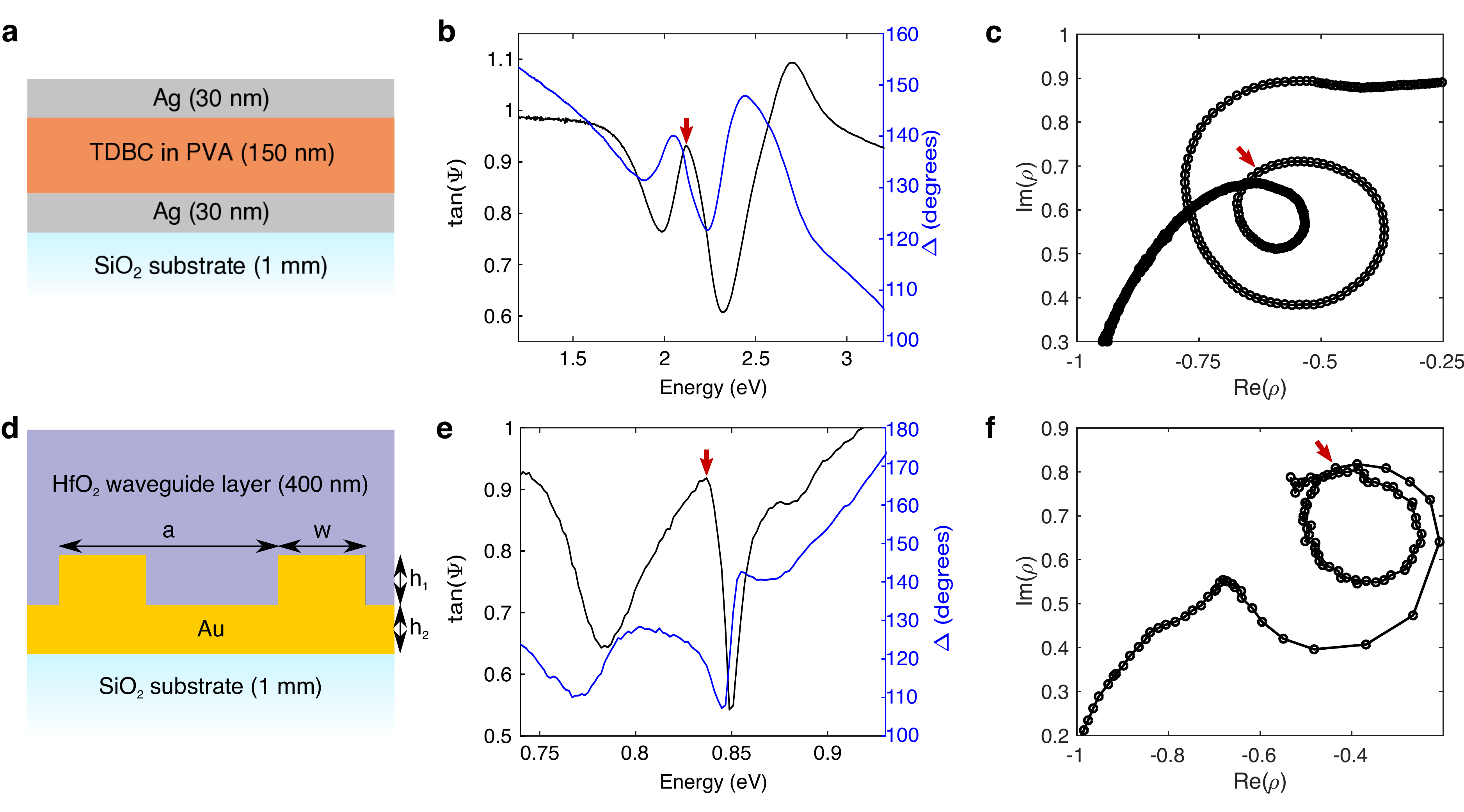}
\centering
\caption{\textbf{Signature of strong coupling in $\rho$ for other systems.} 
The plots of $\tan(\Psi)$/$\Delta$ and $\rho$ for strong coupling in (a-c) a TDBC microcavity and (d-f) a hybrid surface lattice resonance/waveguide structure both feature the same secondary loop observed in figure \ref{fig:rho}. 
$\theta = 60^{\circ}$ for all measurements.
(a) Schematic of the TDBC microcavity measured in (b-c).
(d) Schematic of the hybrid surface lattice resonance/waveguide structure measured in (e-f), where $a=1550$ nm, $w = 450$ nm, $h_1=75$ nm and $h_2 = 65$ nm.
The points corresponding to the maximum value of $\tan(\Psi)$ between the two polariton minima are indicated by red arrows.
}
\label{fig:other}
\end{figure}

Secondary loops in $\rho$ are not unique to MC microcavities.
Figures \ref{fig:other}a-c show strong coupling between an optical microcavity mode and the excitonic resonance (2.1 eV) in the J-aggregate TDBC\cite{dovzhenko2018light} (see figure \ref{fig:other}a for sample design and Supplementary Information for fabrication details).
Figure \ref{fig:other}b shows a splitting of the phase response of the initial cavity mode.
Figure \ref{fig:other}c shows a secondary loop in $\rho$, similar to the one shown in figure \ref{fig:rho}c.

Figures \ref{fig:other}d-f show strong coupling between plasmonic surface lattice resonances and optical waveguide modes.
The structure (figure \ref{fig:other}d) is a one-dimensional gold grating (period 1.55 µm, grating element width 450 nm and height 70 nm) on a 65 nm thick gold sublayer all covered by a 400 nm thick layer of hafnium(IV) oxide (see Supplementary Information for fabrication details).
The grating structure supports plasmonic surface lattice resonances and the hafnium(IV) oxide layer supports guided modes which can become strongly coupled to the plasmonic surface lattice resonance at around 0.8 eV\cite{thomas2017strong}.
The parameter $\rho$ for such a system is plotted in figure \ref{fig:other}f, which also shows a secondary loop. 
The secondary loop varies in size as the incident angle changes (see Supplementary Figure S4).
The innermost points of the secondary loops and their associated values of $\tan(\Psi)$ and $\Delta$ are indicated by red arrows.
In the SPI/MC microcavity (figure \ref{fig:rho}), where the polariton bands have roughly equal amplitude, the innermost point of the secondary loop corresponded to a point between the two polariton bands.
For the TDBC/cavity and SLR/waveguide (figure \ref{fig:other}), where the polariton bands have very different amplitudes, the innermost points of both secondary loops correspond to the minima of the weaker polariton band.

\section{Discussion}

The existing criteria for strong coupling depend variously on the coupling strength $g$, the Rabi splitting $\Omega$, the losses of the confined mode of the electric field and molecular resonator ($\gamma_c$ and $\gamma_m$, respectively) and the energy of the uncoupled cavity mode and molecular excitation $E_{c,m}$.
(For strong coupling $E_{c}\approx E_m \equiv E_0$.)
These criteria are summarised in table \ref{tab:criteria} and discussed in detail in the Supplementary Information.
Here we apply these criteria to our results and compare them with the formation of the secondary loop in figure \ref{fig:rho}.

\begin{figure}
\includegraphics[scale=0.65]{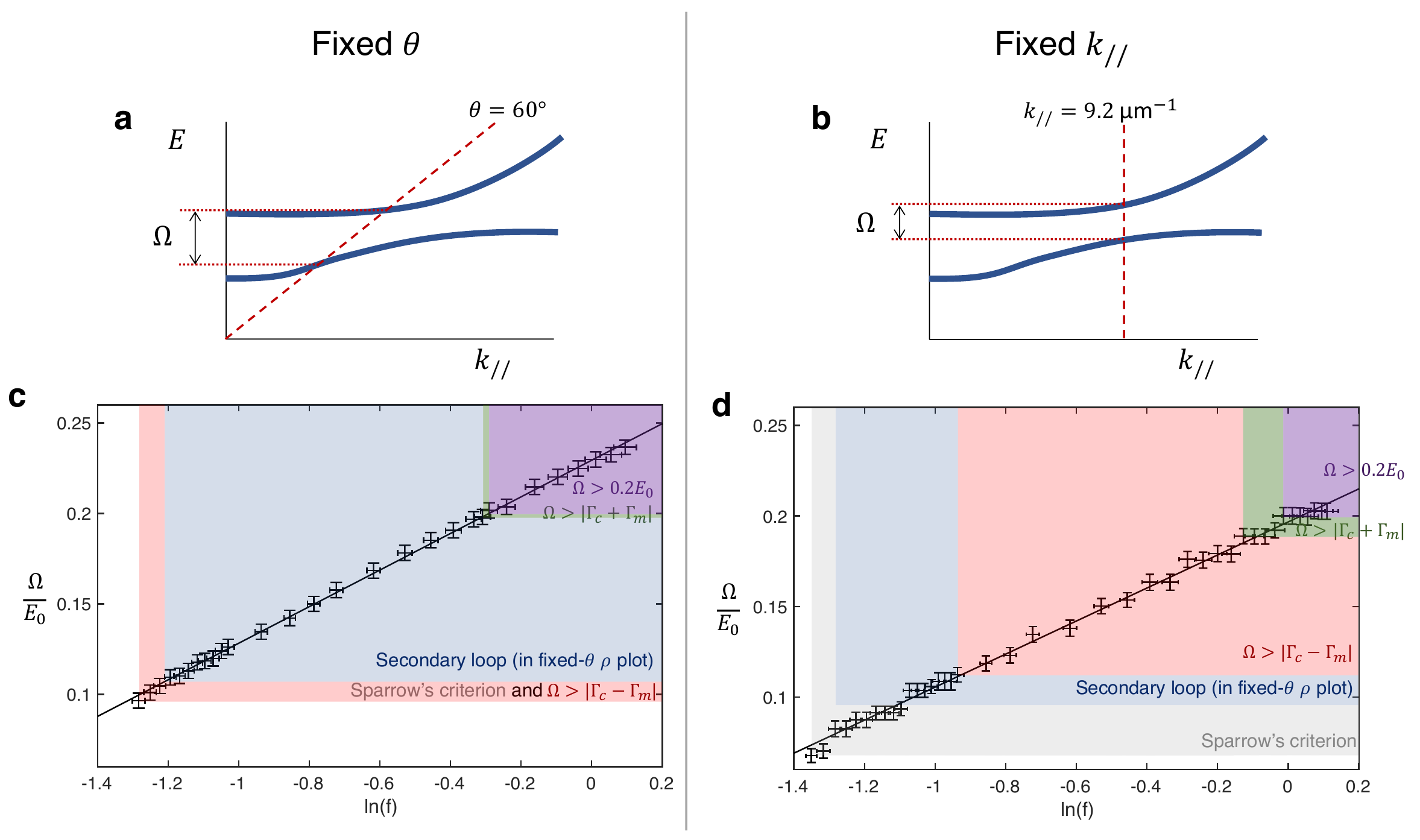}
\centering
\caption{\textbf{Comparison of different criteria for strong coupling.} (a) and (b) show how fixed-angle ($\theta$) and fixed-wavevector ($k_{//}$) spectra are projected on to dispersion plots. (c) and (d) show how the calculated Rabi splitting from (c) fixed-$\theta$ ($\theta=60^{\circ}$) and (d) fixed-wavevector ($k_{//}=9.2$ µm$^{-1}$) spectra changes as a function of Lorentz oscillator amplitude. The shaded regions of these plots show where different strong coupling criteria are fulfilled.}
\label{fig:compare}
\end{figure}

While we cannot directly compare the coupling strength $g$ with the losses $\gamma$, we can compare the experimentally measurable Rabi splitting $\Omega$ with the full-width-half-maxima (FWHM) of the uncoupled resonances of the cavity $\Gamma_{c}$ and the MC molecular transition $\Gamma_{m}$.
We modelled the SPI/MC microcavity using a Fresnel model (see Supplementary Information for details).
The MC resonance was modelled using a single Lorentz oscillator:
\begin{align*}
    \epsilon_{m} = \frac{fB_{m}E_0}{E_0^2 - E^2 - iEB_{m}}.
\end{align*}
$f$ is a dimensionless strength, $B_{m} \approx \Gamma_{c}$ and $E_0$ the energy of the molecular resonance.
The increase in MC molecules was modelled by increasing $f$.
On average the calculated $\Omega$ differed from the experimentally observed $\Omega$ by $8\%$, which is a good level of agreement for such a simple model.
In figure \ref{fig:compare} we plot $\Omega/E_0$ as a function of $f$.
It is possible to measure $\Omega$ and FWHMs in two ways: one is to use a fixed-$\theta$ spectrum (figure \ref{fig:compare}a: this is how all data in figures \ref{fig:time}-\ref{fig:other} were acquired); the other is to use a fixed-$k_{//}$ spectrum (figure \ref{fig:compare}b). 
We plot calculated fixed-$\theta$ $\Omega/E_0$ in figure \ref{fig:compare}c and calculated fixed-$k_{//}$ $\Omega/E_0$ in figure \ref{fig:compare}d.

\begin{table}
\begin{tabular}{|l|l|l|}
\hline
\textbf{Name}        & \textbf{\begin{tabular}[c]{@{}l@{}}Criterion\\ (theory)\end{tabular}} & \textbf{\begin{tabular}[c]{@{}l@{}}Criterion\\ (experiment)\end{tabular}} \\ \hline
Sparrow's\cite{sparrow1916spectroscopic}            & \multicolumn{2}{l|}{\begin{tabular}[c]{@{}l@{}}The spectral midpoint between two \\ resonances shows a local minimum\end{tabular}}                  \\ \hline
Savona \textit{et al.}\cite{savona1995}               & $4g > | \gamma_c - \gamma_m |$                                                           & $\Omega > | \Gamma_c - \Gamma_m |$                                                                  \\ \hline
PT-symmetric Savona \textit{et al.}\cite{savona1995,bender1998}  & $4g >  \gamma_c + \gamma_m $                                                                    & $\Omega >  \Gamma_c + \Gamma_m $                                                                     \\ \hline
Ultrastrong coupling\cite{kockum2019ultrastrong} & $g/E_0 > 0.1$                                                             & $\Omega/E_0 > 0.2$                                                               \\ \hline
\end{tabular}
\caption{\textbf{Criteria for strong coupling.} Summary of different criteria for strong coupling used in the literature.
$g$ is the coupling strength, $\Omega$ is the Rabi splitting, $\gamma_{c,m}$ and $\Gamma_{c,m}$ are the losses and FWHM of the confined electric field mode and molecular resonator, respectively, and $E_0$ is the uncoupled transition energy of the electromagnetic cavity mode and the molecular resonance (which are assumed to be approximately equal).
See Supplementary Information for a detailed discussion of each criterion.}
\label{tab:criteria}
\end{table}

The shaded regions in figures \ref{fig:compare}c and \ref{fig:compare}d show the limits of the various strong coupling criteria described above: Sparrow's criterion (grey), the Savona \textit{et al.} criterion (red), the PT-symmetric Savona \textit{et al.} criterion (green) and the ultrastrong coupling criterion (purple).
We have also plotted the region (shaded blue) in which the secondary loop shown in figure \ref{fig:rho} appears.
The differences between the two plots can be explained by the difference in $\Omega$ (and to a lesser extent by the difference in FWHM) measured in the two configurations.

The relationships between $\Omega/E_0$ and $\ln(f)$ in figures \ref{fig:compare}c and \ref{fig:compare}d are described well by a linear-log plot.
For figure \ref{fig:compare}c:
\begin{align*}
    \Omega/E_0 = 0.10\ln(f) + 0.23;
\end{align*}
and for figure \ref{fig:compare}d:
\begin{align*}
    \Omega/E_0 = 0.09\ln(f) + 0.20.
\end{align*}
In the fixed-$\theta$ case the scan line in figure \ref{fig:compare}a intersects each polariton band at different $k_{//}$, giving a larger $\Omega$ and a larger rate of splitting with increasing $f$.
This explains why the fixed-$\theta$ gradient is $9\%$ higher than the fixed-$k_{//}$ gradient and the $16\%$ difference in y-intercept between the two fits, which is almost identical to the increase in $\Omega$ when moving from fixed-$k_{//}$ to fixed-$\theta$ (on average $17\%$).

The higher values of $\Omega$ in fixed-$\theta$ plots affect the points at which the various criteria for strong coupling are satisfied.
The ultrastrong coupling criterion depends solely on $\Omega/E_0$, so it requires a lower value of $f$ to be fulfilled in fixed-$\theta$ plots.
The Savona \textit{et al.} and PT-symmetric criteria depend on $\Omega$ and $\Gamma_{c,m}$.
As the change in $\Omega$ is much larger than the changes in $\Gamma_{c,m}$ when moving from fixed-$k_{//}$ to fixed-$\theta$, the Savona \textit{et al.} and PT-symmetric criteria are also fulfilled at slightly lower $f$ and $\Omega$.
The higher $\Omega$ value in the fixed-$\theta$ case is sufficiently large that the Savona \textit{et al.} criterion is fulfilled at the same point as Sparrow's criterion (that is, when two resonances are first resolved).
Overall, these differences are relatively small, and the relative stringencies of the criteria are largely unchanged when moving from fixed-$k_{//}$ to fixed-$\theta$ spectra.
A fixed-$\theta$ spectrum can thus provide a similar level of information to that obtained from a fixed-$k_{//}$ spectrum in the analysis of strong coupling experiments.

The most commonly-used criterion (the PT-symmetric Savona \textit{et al.} criterion) is much more stringent, in our case being comparable with the ultrastrong coupling criterion.
The variant of this criterion used in experimental analysis\cite{torma2014strong} ($\Omega > \Gamma_{c} + \Gamma_{m}$) only approximately matches the originally derived criteria since it utilises $\Omega$ and $\Gamma_{c,m}$, not $g$ and $\gamma_{c,m}$.
For this reason we suggest that the PT-symmetric Savona \textit{et al.} criterion is consistently too conservative in defining the transition from weak to strong coupling.

Indeed, using FWHMs in a strong coupling criterion is generally problematic.
In disordered organic molecules such as MC the FWHM can be an unreliable estimate of lifetime.
The FWHM of the absorption peak is often predominantly defined by the vibrational modes within the molecule which split the excitation into many closely spaced modes\cite{lidzey1998strong, houdre1995vacuum, fidder1991optical}.
Furthermore, a criterion for strong coupling that uses FWHMs will be dependent upon the measurement apparatus, not just the system under interrogation.
For oblique angles of incidence the measured FWHMs of modes differ depending upon whether spectra are fixed-$\theta$ or fixed-$k_{//}$ (compare figures \ref{fig:compare}a,b: the measured value of FWHM will depend on how the red line corresponding to the measured spectrum intersects any resonances).
Additionally, it is sometimes simply not possible to characterise the uncoupled modes of a system\cite{brimont2020strong}.
It seems that the most commonplace criteria for strong coupling are somewhat limited since they rely on comparisons of FWHMs and Rabi splitting.

How else can we characterise the transition from weak to strong coupling?
Ideally, a criterion for strong coupling should not be dependent upon the measurement technique.
If a system is in the strong coupling regime this should clearly be apparent in multiple measuring techniques.
Spectroscopic ellipsometry allows one to observe signatures of strong coupling in both amplitude and phase measurements.
The formation of the secondary loop in figure \ref{fig:rho} corresponds to the point at which the amplitude and phase signatures of strong coupling are both observed.
In figure \ref{fig:rho}c the difference between $\Omega$ required for the Savona \textit{et al.} criterion and secondary loop formation is $10\%$.
This is less than the difference ($12\%$) between the experimental and calculated values of $\Omega$ at the point of secondary loop formation, suggesting that the two criteria have a similar level of stringency.
The secondary loop criterion has two advantages over the Savona \textit{et al.} criterion: first, it is not dependent upon the approximation that losses can be equated with FWHM; second, whilst it can be impossible to determine if the Savona \textit{et al.}  criterion has been fulfilled in high-loss systems, observing the secondary loop in ellipsometry (a very low-noise technique) is straightforward.
Furthermore, verifying the existence of a secondary loop in spectroscopic ellipsometry requires one to take just one measurement at a single angle.
For these reasons, we suggest that studying the ellipsometric topology of a system and observing a secondary loop in $\rho$ could perhaps provide an alternative and useful criterion for strong coupling.

\section{Conclusions}

We have studied the transition from the weak to strong coupling regime in a MC microcavity using spectroscopic ellipsometry and observed a signature for strong coupling in the ellipsometric phase response.
Combining amplitude and phase data produces a topologically distinct feature that we associate with strong coupling.
The observation of this feature for strong coupling of both molecular/microcavity and surface lattice resonance/guided mode structures suggests it is a more general signature of strong coupling.
We have compared the emergence of this change in ellipsometric topology with existing criteria for strong coupling and suggest that ellipsometric topology could provide an alternative and useful criterion for strong coupling.
In summary, our results suggest a new criterion for strong coupling that does not suffer from the limitations of existing strong coupling criteria.
More widely, our results suggest that spectroscopic ellipsometry may provide a powerful probe with which to explore strong coupling.

\section*{Supporting Information}
Fourier transmission spectroscopy schematic; MC transmittance spectrum; ellipsometry of uncoupled systems; strong coupling of surface lattice resonances and waveguide modes; review of strong coupling criteria.
This material is available free of charge via the internet at https://pubs.acs.org/



\section*{Competing interests}
The authors declare no competing interests.

\newpage
\begin{figure}
\includegraphics[scale=0.5]{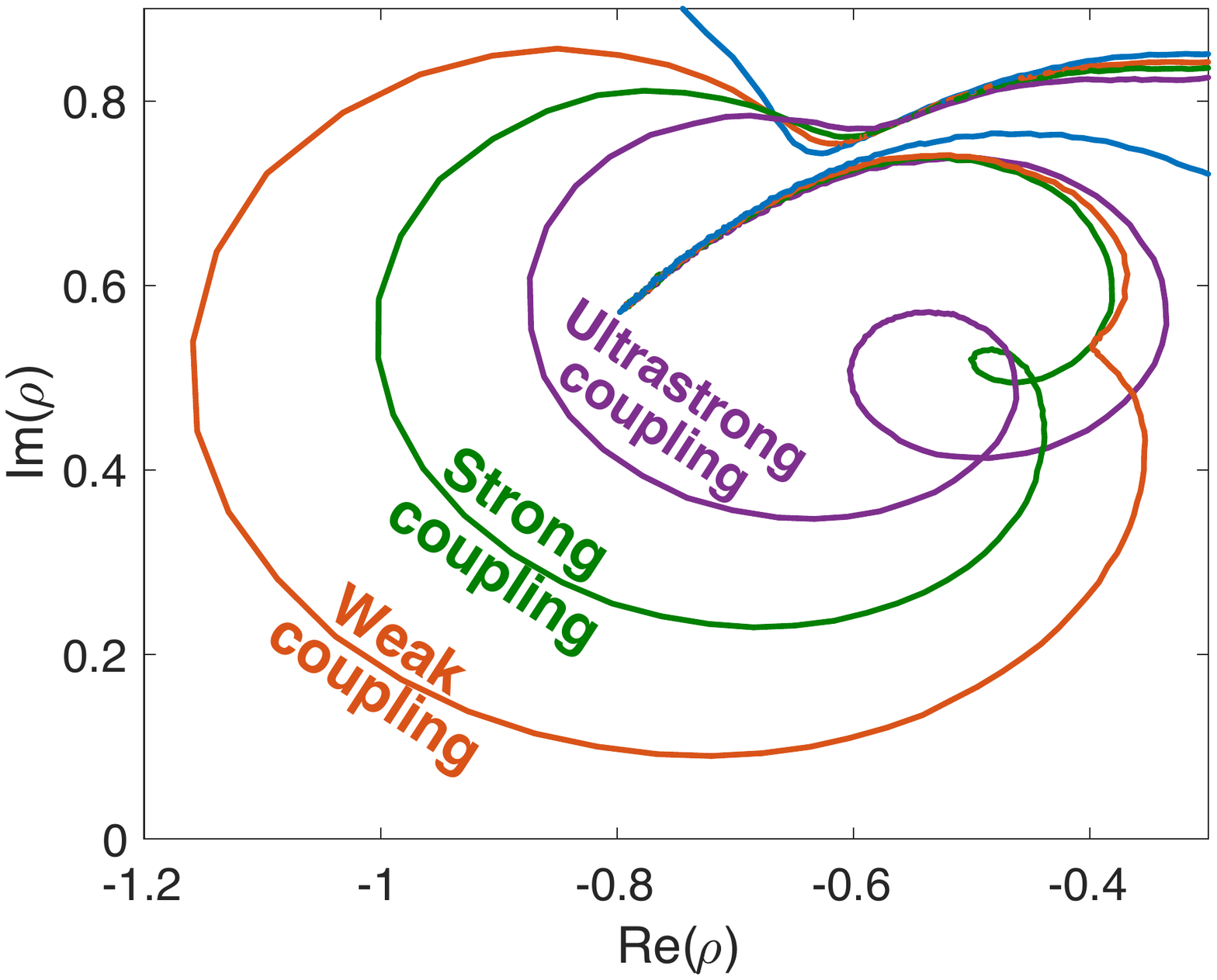}
\centering
\caption{TOC image}
\end{figure}

\end{document}


\maketitle

\newpage
\section{Methods}
\subsection{Fabrication of optical microcavities}
The microcavity samples used in this study consist of a layer of polymer matrix sandwiched between two thin Ag films. 
One sample is filled with photochromic spiropyran (SPI) molecules (1', 3'-dihydro-1', 3', 3'-trimethyl-6-nitrospiro[2H-1-benzopyran-2, 2'-(2H)- indole]) in a PMMA (polymethyl methacrylate) matrix, while the other sample is filled with polyvinyl alcohol (PVA) doped with TDBC molecules (5,6-dichloro-2-[[5,6-dichloro-1-ethyl-3-(4-sulphobutyl)-benzimidazol2-ylidene]-propenyl]-1-ethyl-3-(4-sulphobutyl)-benzimidazolium hydroxide, sodium salt, inner salt).

First, a Ag film (thickness 30 nm) is deposited on a glass slide (thickness 1 mm) by thermal evaporation at a rate of 0.1 nm/s. 
The polymer matrix is then spin-coated on top of the silver film. 
Finally, the upper Ag layer is deposited on top of the polymer matrix using the same parameters as for the first Ag film.

The SPI/PMMA matrix was prepared by dissolving PMMA (molar weight 996 000) in toluene (1 wt$\%$ PMMA). 
SPI is then dissolved in the PMMA-toluene solution with a weight ratio of 3:2 SPI to PMMA. 
The solution is then filtered using a syringe filter of pore size 0.2 µm. 
A spin speed of 3000 rpm was used to deposit the SPI/PMMA matrix to achieve a thickness of 150 nm.

The TDBC/PVA matrix was prepared by mixing PVA (molar weight 85 000 – 124 000) in water at 90 $^{\circ}$C for several hours to obtain a solution of 2.3 wt$\%$. 
Once the PVA had completely dissolved the solution was left to cool down before adding TDBC (0.8 wt$\%$). 
A syringe filter of pore size 0.5 µm was used to filter the solution. 
The TDBC/PVA matrix was spin coated using a spin speed of 2000 rpm to achieve a thickness of 150 nm, suitable for a first-order cavity resonance to match the TDBC exciton.

\subsection{Fabrication of hybrid plasmonic waveguide structure}

First, a Cr (3 nm-thick adhesion layer) and Au (thickness 65 nm) film were deposited on a glass substrate (thickness 1 mm) using electron beam evaporation. 
Then the plasmonic grating structure (total area 300 µm $\times$ 100 µm, grating period 1.55 µm, grating element width 450 nm and height 70 nm) was fabricated using electron beam lithography and electron beam evaporation. 
Finally, a hafnium oxide layer (thickness 400 nm) was deposited on top of the structure using electron beam evaporation.

\subsection{Spectroscopic ellipsometry}
Spectroscopic ellipsometry was carried out using a J. A. Woollam Co. M-2000XI with which we measured the ellipsometric parameters $\Psi$ and $\Delta$ in the wavelength range 210-1690 nm, with a wavelength step of 1.5 nm for 210-1000 nm and 3.5 nm for 1000-1690 nm. 
$\Psi$ gives the ratio of the field reflection coefficients for $p$- and $s$-polarised light (the moduli of $r_p$ and $r_s$, the complex Fresnel reflection coefficients for $p$- and $s$-polarised light  respectively) and $\Delta$ is the phase difference between the same coefficients such that $r_p/r_s = \tan(\Psi) e^{i\Delta}$. 
Since ellipsometry measures the ratio of two signals it cancels out a lot of noise from the source, making it a very sensitive measuring technique. 
The light source in the M-2000XI was a 75 W Xe arc lamp which produced a a smooth ultraviolet continuum\cite{baum1950ultraviolet} that was used to convert SPI to MC.
For consistency all measurements were taken at an incident angle of $60^{\circ}$.

\subsection{Modelling of SPI/MC microcavity}
We modelled the SPI/MC microcavity using CompleteEASE\textregistered\cite{completeease}.
The structure was SiO$_2$ substrate/Ag/ dielectric/Ag.
Standard literature values were used for the SiO$_2$ and Ag optical constants.
The permittivity of the dielectric host containing SPI/MC, $\epsilon_{\text{SPI/MC}}$, was modelled  using a single Lorentz oscillator:
\begin{align*}
    \epsilon_{\text{SPI/MC}} = \epsilon_b + \frac{fB_{m}E_0}{E_0^2 - E^2 - iEB_{m}},
\end{align*}
where $\epsilon_b = 2.25$ is the background permittivity, $f$ is the dimensionless amplitude, $B_{m}$ is approximately the FWHM and $E_0$ the energy of the molecular resonance.
$f$ was allowed to vary to model the increase in MC molecules with the progression of time.
To improve the quality of fit at earlier measurements $B_{m}$ and $E_0$ were set as fit parameters.

\newpage
\section*{Figure S1: Fourier transmission spectroscopy}
\begin{figure}[!h]
\includegraphics[scale=1.00]{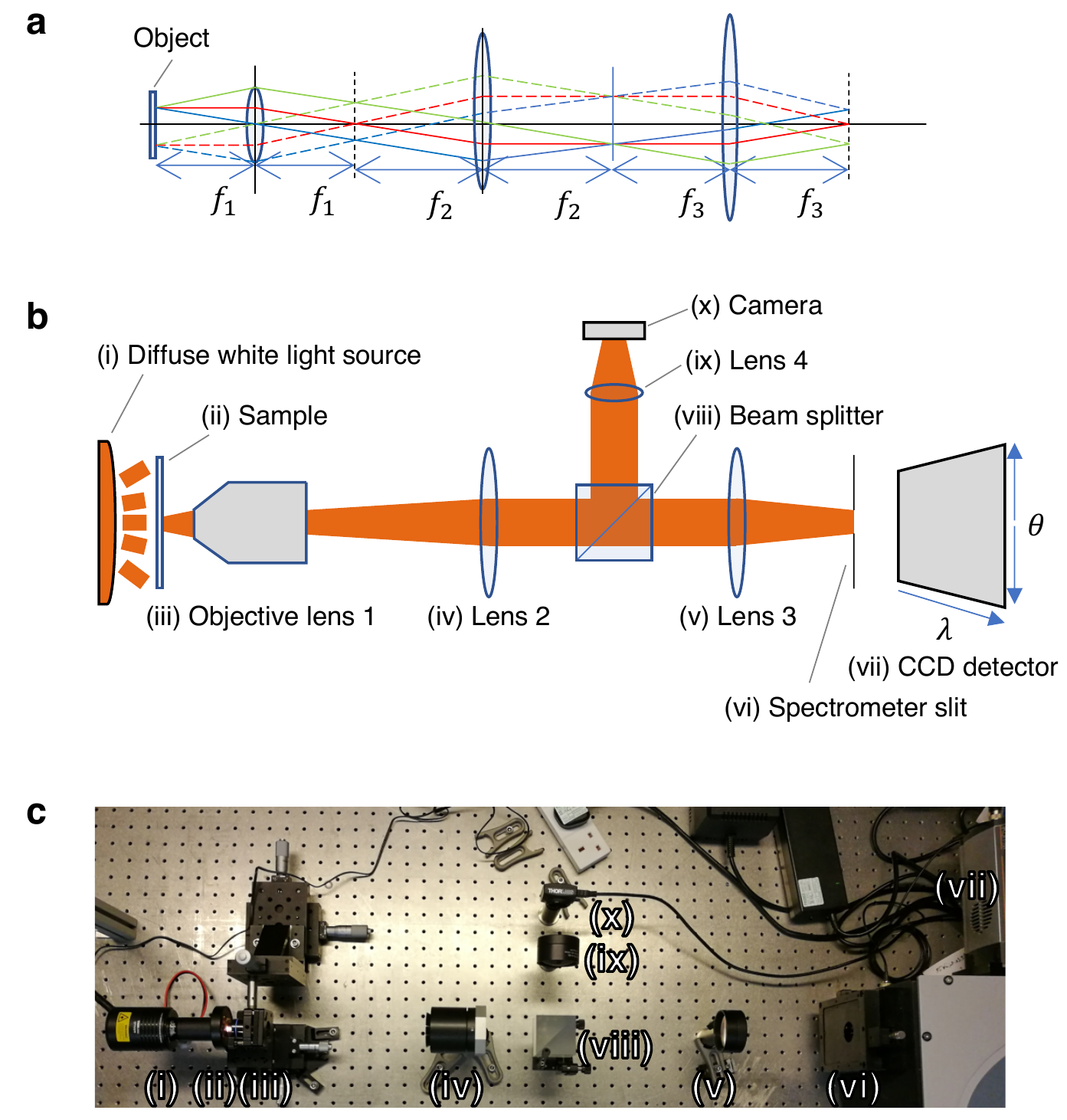}
\centering
\end{figure}
\noindent (a) General principle of Fourier plane imaging: light from the object on the left passes through three lenses.
Real images of the object form in the planes indicated by vertical solid lines; the Fourier transform of the real image exists in the planes indicated by vertical dashed lines.
\newline \newline
(b) Schematic of Fourier transmission spectroscopy experiment: the Fourier plane after the third lens forms on the entrance slit to a spectrometer.
The spectrometer's output is captured by a CCD detector.
Each column of CCD pixels corresponds to a different wavelength of light; each row of pixels corresponds to light transmitted (or emitted) from the sample at a different angle of incidence.
The CCD image is processed to give a dispersion plot such as the one shown in figure 1a of the main manuscript.
A beam splitter, lens and camera (elements (viii)-(x)) can be added to the experiment to image the sample.
The light path, shown in orange, is indicative only.
\newline \newline
(c) Top-down photograph of our Fourier transmission experiment.
Our light source (i) is a quartz tungsten-halogen lamp with aspheric lenses and diffusers; our lenses have effective focal lengths of (iii) 4.5 mm (objective lens, numerical aperture 0.65), (iv) 180 mm, (v) 200 mm and (ix) 75 mm.

\newpage
\section*{Figure S2: MC transmittance spectrum}
\begin{figure}[!h]
\includegraphics[scale=0.50]{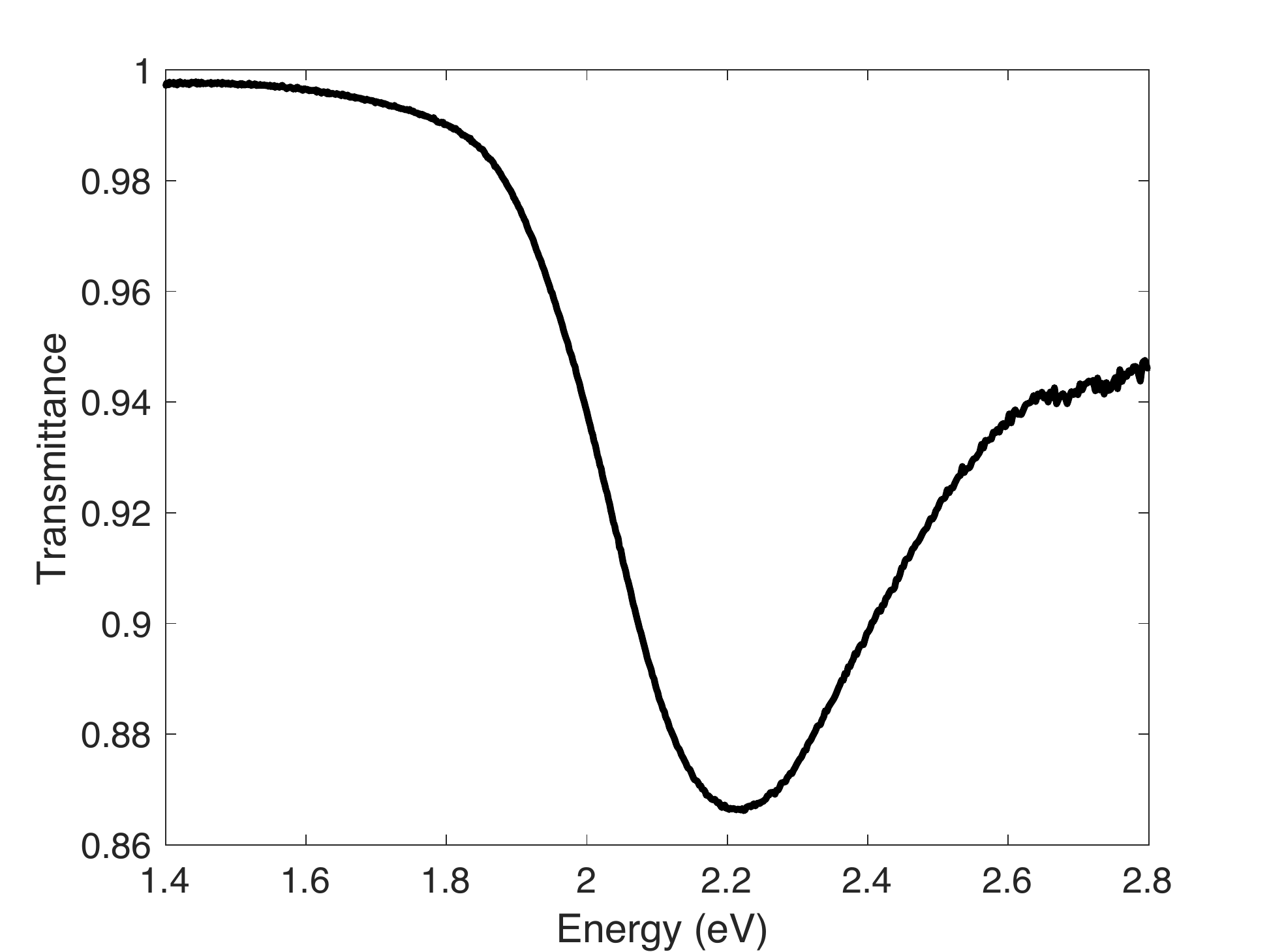}
\centering
\end{figure}
\noindent Transmittance through a merocyanine (MC) film (thickness 150 nm) spin-coated on a glass substrate, normalised against transmission for an uncoated substrate (see Methods for fabrication).

\newpage
\section*{Figure S3: Ellipsometry of uncoupled systems}
\begin{figure}[!h]
\includegraphics[scale=1.0]{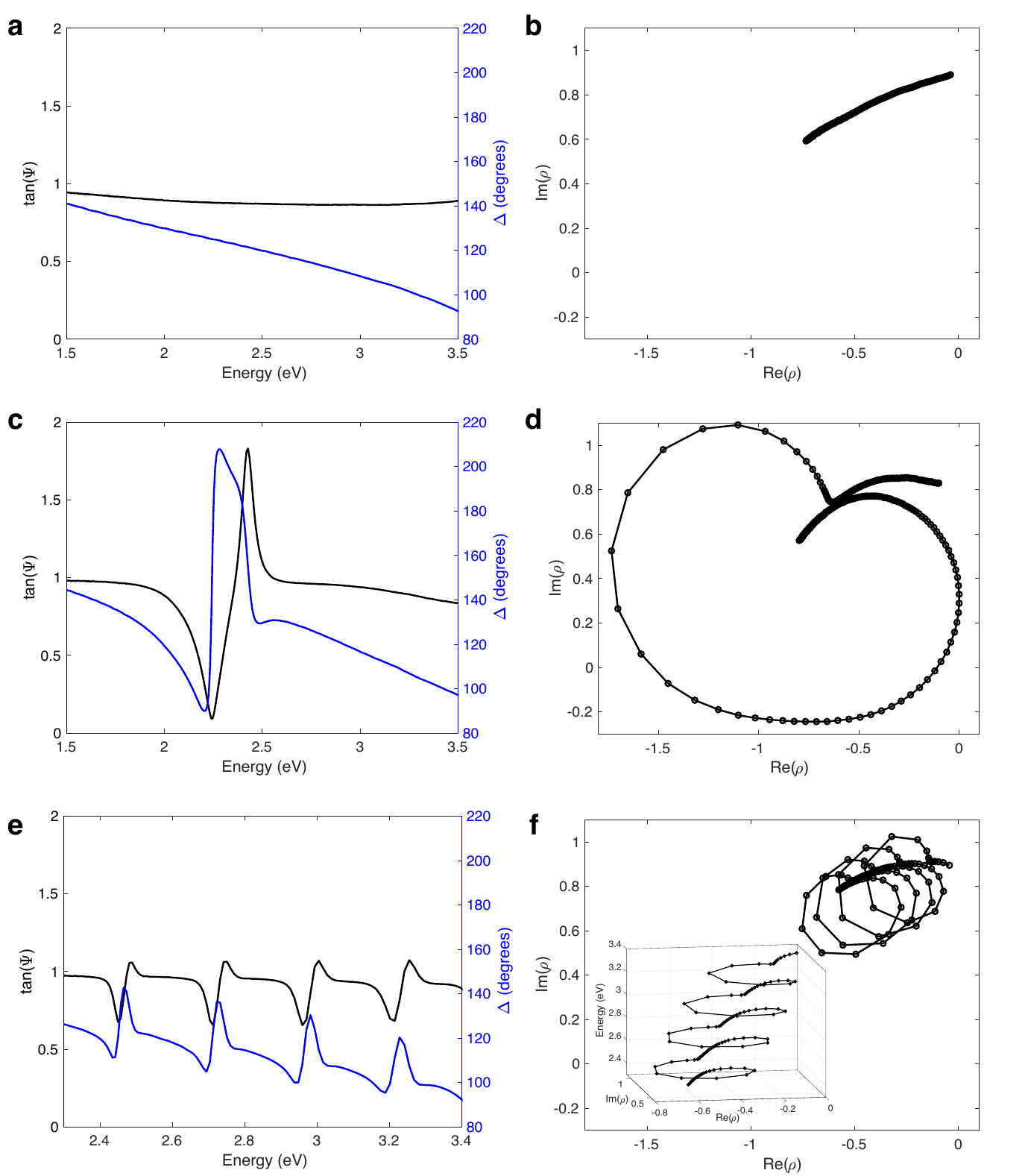}
\centering
\end{figure}

\noindent The ellipsometric parameters $\tan(\Psi)$ and $\Delta$ (a, c, e) and $\rho$ (b, d, f) (all measured at $\theta=60^{\circ}$) for:

\begin{itemize}
    \item (a, b) a thin film of Ag (thickness 40 nm) on a glass substrate.
    (b) nicely shows the Ag arc mentioned in the main text;
    \item (c, d) the Ag (30 nm)/SPI in PMMA (150 nm)/ Ag (30 nm) microcavity structure (Figure 1b)  supporting one uncoupled first-order cavity mode before exposure to UV radiation. Notice how the Ag arc in (b) has been modified to include a loop due to the cavity mode;
    \item (e, f) a multimode Ag/PMMA/Ag cavity (PMMA thickness $\sim 2$ µm) supporting multiple uncoupled cavity modes.
    The loops in $\rho$ corresponding to the uncoupled cavity modes overlap one another (f).
    This should be contrasted with the coupled resonances in Figures 3-4 where the two loops merge to form a secondary loop.
    Inset: 3D plot of $\rho$ with energy on the z-axis showing that each uncoupled cavity mode is associated with a separate loop.
\end{itemize}

\newpage
\section*{Figure S4: Strong coupling of surface lattice resonances and waveguide modes}
\begin{figure}[!h]
\includegraphics[scale=0.6]{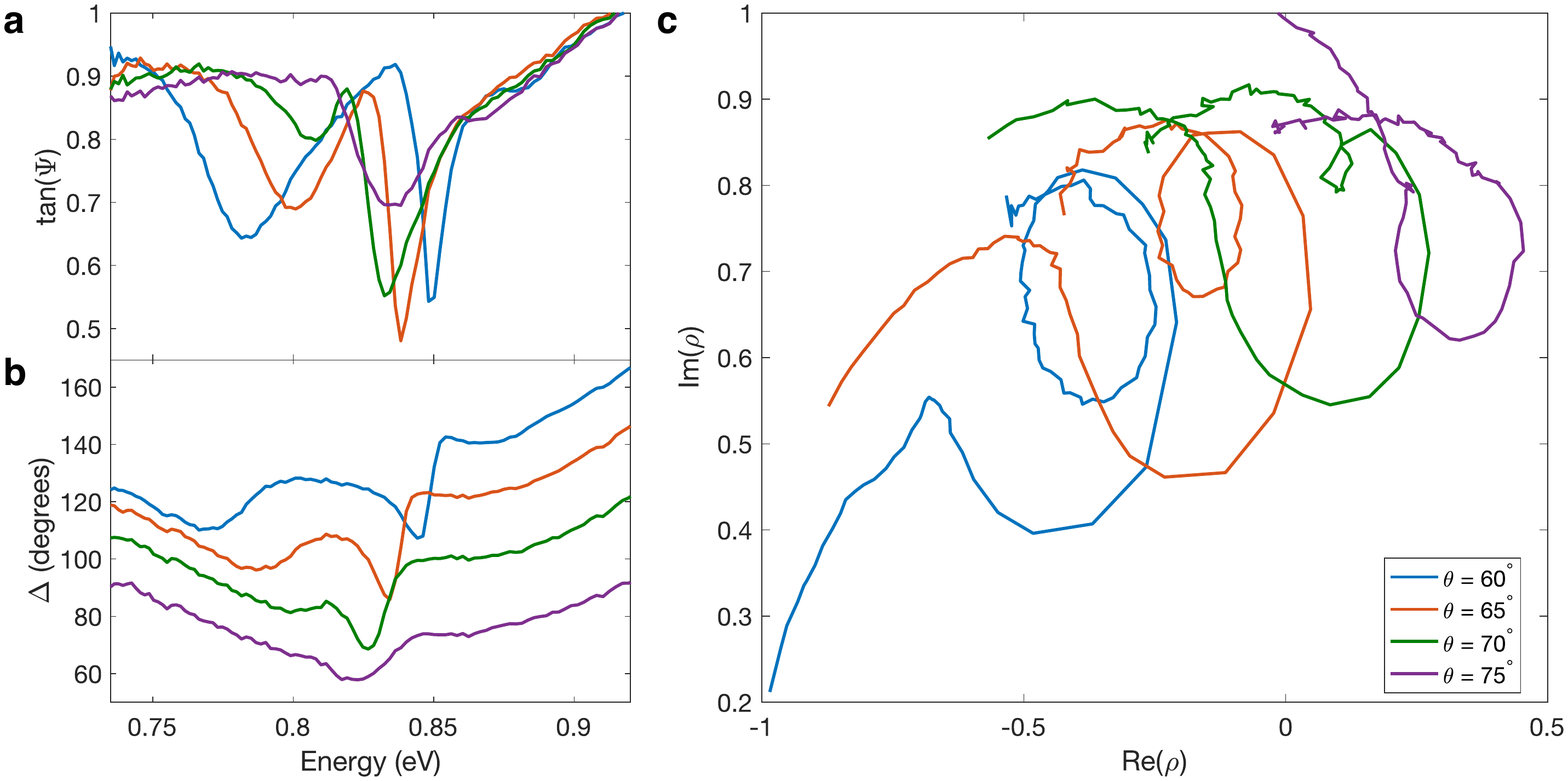}
\centering
\end{figure}

\noindent  (a), (b) and (c) show the measured
ellipsometric parameters $\tan(\Psi)$, $\Delta$ and $\rho$, respectively, for the SLR/waveguide structure described in figure 4d-f measured at incident angles of $60^{\circ}$ (blue curve), $65^{\circ}$ (orange curve), $70^{\circ}$ (green curve) and $75^{\circ}$ (purple curve). The secondary loop grows in size as the angle of incidence is decreased and the splitting of the polariton branches increases. The similarity of this plot to Figure 3 shows that the secondary loop is not unique to planar microcavity systems, suggesting that the use of the secondary loop as a strong coupling criterion is applicable to a wider selection of strongly coupled systems.

\newpage

\section{Criteria for strong coupling}

The existing criteria for strong coupling depend on a variety of parameters such as the coupling strength $g$, the Rabi splitting $\Omega$, the losses of the confined electric field and molecular resonator ($\gamma_c$ and $\gamma_m$, respectively) and the energy of the uncoupled modes $E_{c,m}$.

\subsection{Intermediate coupling}

Arguably the simplest criterion for strong coupling is to look for the point at which one peak splits into two.
This point is defined by Sparrow's criterion: two maxima in a spectrum are considered resolved if the midpoint between them shows a minimum in intensity\cite{sparrow1916spectroscopic}.
This criterion, however, can be misleading. 
It is possible to misinterpret apparently split spectral features as being Rabi split even when the system is outside the strong coupling regime\cite{zengin2016evaluating, stete2017signatures}.
For example, so-called induced transparency can produce a dip in a resonance caused by Fano interference between a quantum dot and plasmonic resonance that can occur entirely in the weak coupling regime\cite{wu2010quantum}.
This has resulted in the classification of an ``intermediate coupling'' regime observed in strong coupling scattering experiments\cite{leng2018strong}.

\subsection{Savona \textit{et al.} criterion}

Savona \textit{et al.}\cite{savona1995} derived a criterion for strong coupling based on the following coupling matrix:
\begin{equation}
\begin{aligned}
\mathcal{H} & =
\left( \begin{array}{cc}
E_c -i\gamma_c/2  & g \\
g & E_{m}-i\gamma_{m}/2
\end{array} \right).
\end{aligned}
\label{eq:savherm}
\end{equation}
The eigenvalues of the coupling matrix $\mathcal{H}$ are
\begin{align*}
    E_\pm = & \frac{1}{2} (E_c + E_m) - i\left(\frac{\gamma_c}{2} + \frac{\gamma_m}{2}\right)  \\ &  \pm \sqrt{4g^2 - (E_c - E_m)^2 - \left(\frac{\gamma_c}{2} - \frac{\gamma_m}{2}\right)^2 - 2i(E_c - E_m)\left(\frac{\gamma_c}{2} - \frac{\gamma_m}{2}\right)}
\end{align*}
and at resonance between cavity photons and excitons (i.e. $E_c = E_m \equiv E_0$), the eigenvalues become
\begin{align}
    E_\pm = E_0 - \frac{1}{2}i\left(\frac{\gamma_c}{2} + \frac{\gamma_m}{2}\right) \pm \frac{1}{2}\sqrt{4g^2 - \left(\frac{\gamma_c}{2} - \frac{\gamma_m}{2}\right)^2}.
    \label{eq:eig_h1}
\end{align}
The eigenvalues and eigenvectors of the coupling matrix give the energies $E_\pm$ and Hopfield coefficients of each polariton branch, respectively.
The coupling matrix in equation \ref{eq:savherm} is non-Hermitian and has broken PT-symmetry \cite{bender1998}: its eigenvalues have a complex component which is independent of the coupling factor $g$.
The splitting $\Omega$ of the upper and lower polariton modes is given by
\begin{align}
    \Omega = E_+ - E_- = \sqrt{4g^2 - \left(\frac{\gamma_c}{2} - \frac{\gamma_m}{2}\right)^2}.
\end{align}
Savona \textit{et al.} argue that the transition from the weak to strong coupling regime occurs when $\Omega$ becomes real, that is, when
\begin{align}
    4g > |\gamma_c - \gamma_m|.
    \label{eq:savona}
\end{align}

This criterion, then, does not compare the experimentally observed Rabi splitting with the linewidths of the uncoupled resonances.
It compares the coupling strength with the \textit{difference} between the loss rates of the two uncoupled resonances.
Equation \ref{eq:savona} implies that a system can be in the strong coupling regime even if the widths of the uncoupled resonances far exceed the splitting - provided the uncoupled resonances have very similar widths.
This suggests that a system can be in the strong coupling regime even if the linewidths of the polariton bands are high enough to obscure anti-crossing.
If equation \ref{eq:savona} is a valid criterion of strong coupling then intensity measurements alone may not be enough to observe all cases of strong coupling.

\subsection{Imposing PT-symmetry}

PT-symmetry could be imposed on equation \ref{eq:savherm} if $\gamma_c = -\gamma_m$.
However, since two identical energy loss factors are never attained in a physical system, we can instead introduce the average energy loss factor $\gamma_{av} = \frac{1}{2}(\gamma_c + \gamma_m)$, which allows us to write the following PT-symmetric coupling matrix:
\begin{equation}
\begin{aligned}
\mathcal{H} & =
\left( \begin{array}{cc}
E_c +i\gamma_{av}/2  & g \\
g & E_{m}-i\gamma_{av}/2
\end{array} \right).
\end{aligned}
\label{eq:ptham}
\end{equation}
At resonance the eigenvalues of $\mathcal{H}$ are
\begin{align}
    E_\pm = E_0 \pm \sqrt{g^2 - \left(\frac{\gamma_{av}}{2}\right)^2}.
\end{align}
In this case the mode splitting (expressed now in terms of $\gamma_c$ and $\gamma_m$) is
\begin{align}
    \Omega = \sqrt{4g^2 - \left(\frac{\gamma_c}{2} + \frac{\gamma_m}{2}\right)^2}.
\end{align}
If $\Omega$ has positive, real solutions in the strong coupling regime then the criterion for strong coupling is
\begin{align}
    4g > \gamma_c + \gamma_m.
\end{align}

This criterion is perhaps the most intuitive when performing analysis of experimental data\cite{khitrova2006vacuum,torma2014strong}.
The mode splitting $\Omega$ is related to the peak splitting in absorption spectroscopy measurements (and transmission/reflection measurements), where the energy loss factors are related to the full-width-half-maxima of the absorption peaks of uncoupled excitons and cavity photons.
Comparing the peak splitting with the average of the peak widths is usually interpreted as a resolution criterion: if the peak splitting is larger than the average of the uncoupled peak widths, then the splitting is measurable, and the system is in the strong coupling regime.
Otherwise, the splitting is not measurable and the system is in the weak coupling regime.

\subsection{Ultrastrong coupling}

Many models exist that can describe strongly coupled systems.
The rotating wave approximation, which neglects certain higher-order perturbative terms, underpins both the Jaynes-Cummings model\cite{jaynes1963comparison} (used to model single-atom systems) and the Tavis-Cummings model\cite{tavis1968exact} (many-atom systems).
When the coupling strength $g$ becomes comparable to the transition energy $E_0$ the higher-order perturbative terms neglected in the rotating wave approximation become significant and different models must be used\cite{kockum2019ultrastrong}.
In these circumstances the system is said to be in the \textit{ultrastrong} coupling regime\cite{ciuti2005quantum}.
A related criterion (known sometimes as the deep strong coupling regime) occurs when the coupling interaction is sufficient to alter the ground state energy of the system\cite{casanova2010deep, yoshihara2017superconducting, forn2017ultrastrong}.
In this regime the higher-order perturbative processes can become not just observable but dominant.

A commonly-used criterion for ultrastrong coupling is\cite{kockum2019ultrastrong,forn2019ultrastrong}
\begin{align}
    \frac{g}{E_0} > 0.1,
\end{align}
but this choice is arbitrary and has no deep physical meaning.
Indeed, the first work recognised as probing the ultrastrong coupling regime achieved $\frac{g}{E_0} = 0.05$\cite{dini2003microcavity}.
In contrast with the previously discussed criteria for strong coupling, this criterion for ultrastrong coupling compares the coupling strength of the system not to the linewidths of uncoupled resonances but to the ground state energy.
Therefore, it is perfectly possible for coupled resonances to fulfil the criterion for ultrastrong coupling criterion but fail to meet any of the aforementioned conventional strong coupling criteria\cite{de2017virtual}.
For these reasons it could be argued that the practical use in defining an ultrastrong coupling regime is not to define a fundamentally different region in phase space but instead to justify the use of the rotating wave approximation when modelling a system.

\newpage
